# How flexible do we need to be? Using electricity systems models to identify optimal designs for flexible carbon capture storage system for gas-fired power plants

Fangwei Cheng[1], Qian Luo[1] and Jesse Jenkins[1,2*]
[1]Andlinger Center for energy and environment, Princeton University
[2]Department of Mechanical and Aerospace Engineering, Princeton University
*Corresponding author (jessejenkins@princeton.edu)

## Abstract

As the share of variable renewable energy in power systems grows, enhancing the operational flexibility of combined cycle gas turbines with carbon capture and storage (CCGT-CCS) becomes increasingly valuable. This study integrates techno-economic analysis with capacity expansion modeling to quantify the value of improved CCGT-CCS flexibility—such as lower start-up costs, reduced minimum generation, faster ramping, and shorter up/down times—at both plant and system levels. Using the Texas power system as a case study, we find that increased flexibility raises CCGT-CCS generation profits and installed capacity. Under various policy scenarios, CCGT-CCS benefits most from a $CO_2$ tax (or equivalent emissions cap), more so than from clean energy standards or capture subsidies like the federal 45Q tax credit. However, electricity system cost savings remain modest, reducing total costs by only 0.3–0.5%. Thus, flexibility improvements should be pursued only if they entail limited increases in capital and maintenance costs.

**Key words:** carbon capture and storage, operational flexibility, techno-economic analysis, capacity expansion models

**Synopsis:** This study demonstrates how improved flexibility of carbon capture-equipped gas plants can impact plant operations, operating profit, and power system costs.

**Introduction**

The market share of wind and solar photovoltaic (PV) generators is increasing in many power systems, yet the weather-dependent nature of these zero emissions generators create challenges in balancing electricity supply and demand. With increasing variable renewable energy (VRE) penetration levels, firm low-carbon generators are expected to play an important role in maintaining the power generation reliability in a carbon constrained electricity grid.[1,2] Combined cycle gas turbines with carbon capture and storage (CCGT-CCS) are one such option. CCGTs are designed for flexible operation and able to increase or decrease output or "ramp" rapidly and cycle on/off within a few hours or less.[3] In contrast, systems to capture $CO_2$ from flue gases were designed primarily for industrial processes that operate constantly or coal-fired power plants assumed to operate in a consistent, "baseload" manner. As a result, CCS systems are not designed for flexible operation and may severely constrain the operational flexibility of CCGT-CCS plants.[4] Innovations in systems design and engineering may therefore be required to increase the flexibility of CCGT-CCS plants and satisfy future grid needs.

The flexibility of CCGT-CCS plants includes technical and economic aspects. The technical aspects include the capability to maintain a lower minimum stable generation level, ramp up and down quickly, and greater start-up and shut-down flexibility (e.g., a lower minimum up and down time)[5]. The economic aspects include lower startup costs, ramping costs, and improved efficiency when operating below nominal capacity (e.g., part load efficiency)[6]. To establish design criteria and innovation priorities for CCGT-CCS plants, it is necessary to understand their potential operating patterns and flexibility requirements in the future electricity grid under relevant policies contexts. The dispatch of CCGT-CCS will be determined by the combination of technical constraints, startup costs, fuel costs, relevant policy requirements, and the economic optimization of generation, storage, and demand-side resources across the grid region.

So far, modeling studies have explored operating characteristics of CCGT-CCS [7,8], most of which use exogenous electricity price series to determine the net-present value (NPV) of CCGT-CCS with varying system designs and performance characteristics. Very few studies have been performed to co-optimize CCGT-CCS operations at the power system level and explore the economic feasibility of operating CCGT-CCS generators flexibly[9]. To fill this gap, this work seeks to provide insights on the value of flexible operation of CCGT-CCS in future electricity systems with increasing VRE penetration levels under relevant policies. Our analysis quantifies the value of flexibility from two aspects. From the plant owner's perspective, we explore how improved flexibility might benefit the plant owner by determining the operating profits of CCGT-CCS under various scenarios. From the public sector's perspective, we estimate the benefits of flexible CCGT-CCS to the electricity system by using the total system cost (TSC) as an indicator of the total cost of electricity supply. We focus on the impacts of carbon policies (i.e. $CO_2$ tax and $CO_2$ subsidy) on the operations and adoption of CCGT-CCS in this work, as they have profound impacts on the relative competitiveness of CCGT-CCS and other competing generators and the position of CCGT-CCS in the economic dispatch of power system assets: under a $CO_2$ tax, generators with higher emissions become more costly, while a $CO_2$ subsidy reduces the costs of generators in proportion to the quantity of $CO_2$ captured per unit of electricity output.

We use GenX[10], an open-source capacity expansion model with unit commitment and economic dispatch to investigate plant-level CCGT-CCS operating patterns, operating profits, and contribution to the power grid under various operational characteristics and policy assumptions. By doing this, we seek to answer the following three questions: (1) What are the potential operating patterns of CCGT-CCS in the future electricity grid with varying VRE penetration levels; (2) How do different carbon policies affect the deployment, operation, and technical flexibility requirements for CCGT-CCS; and (3) How do improved flexibility characteristics affect CCGT-CCS capacity deployment and power system costs.

**Methods**

This study employs GenX, a highly configurable open-source electricity system capacity expansion model to determine the optimal investments and operating decisions to meet future electricity demand at least cost, while subject to a set of engineering, economic, and policy constraints. The input data was prepared by PowerGenome[11], an open-source software designed to provide input files for electricity capacity expansion model. The input data includes capacity of existing generators, the cost (e.g., capital cost, fixed and variable operating and maintenance cost, startup costs) and heat rate assumptions of existing and new-build resource options, hourly demand profile, wind and solar variability profiles, etc. The cost assumptions associated with new-build resources were from NREL Annual Technology Baseline (ATB) 2022[12], where we used moderate cost assumptions for all the new generators (see Table S1 for details). NREL ATB provides cost assumptions based on their projections for each year; we therefore use the average costs over the planning period (2023-2035). The fuel prices are derived from EIA AEO 2022[13], and we used natural gas prices of $2.8/MMBTU for all the runs.

We use the Texas power system (ERCOT) as a study region due to its isolation from the larger national grid, high renewable energy potential and natural gas supply, and ideal locations for $CO_2$ storage. We implement relevant policies established by the Inflation Reduction Act of 2022, including production tax credits and investment tax credits for new wind, solar, battery, and nuclear[14]. We also purposely prevent nuclear generators from economic retirement to reflect the impact of the production tax credit for existing reactors and related policy support[15]. We first run GenX to determine the optimal mix of capacity and operations for ERCOT to meet estimated electricity demand in 2035 at minimum cost (Figure S1). New resources considered in the capacity expansion modeling include wind, solar PV, battery energy storage, hydrogen combustion turbine (CT) and combined cycle gas turbine (CCGT), natural gas CT and CCGT, and nuclear. For each scenario, we first run capacity expansion without CCGT-CCS. Next, we fix the ending capacity of other resources, add a 500 MW CCGT-CCS plant to this system, and perform unit commitment and economic dispatch runs to explore the operation paradigms (e.g., start up and shut down events, capacity factor, dispatch profiles, etc.) and economic performance (e.g., operating profits) of CCGT-CCS plants with varying flexibility characteristics. Last, we consider cases where CCGT-CCS is included in the capacity expansion to assess the optimal capacity penetration and value of flexible CCGT-CCS for the power system by comparing the total power system cost with and without improved CCGT-CCS flexibility.

Given the nascency of the technology and variety of possible system designs, the precise operating constraints for CCGT-CCS remain uncertain at this time. However, we anticipate CCGT-CCS will be more flexible than a pulverized coal plant with CCS but less flexible than

CCGT without CCS. Thus, we use the technical flexibility parameters from coal CCS (mostly derived from the Petra Nova demonstration facility[22], except for ramping rates, which are derived from a CCGT-CCS demo project[19]) and conventional CCGT without CCS to represent the most inflexible and flexible cases for CCGT-CCS, respectively (see Table 1 for detailed descriptions). In summary, the most flexible/inflexible CCGT-CCS has minimum stable power generation of 30%/70% of nominal capacity, ramping rates of 100%/36% of nominal capacity per hour, minimum up time of 4hr/12hr after start-up, and minimum down time of 4hr/18hr after shut-down. We also assume the most flexible/inflexible CCGT-CCS has 100%/150% of the fixed start-up costs of a conventional CCGT. The startup fuel consumption is assumed to be the same. We exclude ramping costs as they are negligible compared to startup costs.[16,17]

**Table 1.** Assumption for the worst and best flexibility characteristics for CCGT-CCS

|  | Ramping rate (%$P_{nom}$/hr) [a] | Minimum up time (hr) [b] | Minimum down time (hr)[b] | Turndown – minimum load (%) [c] | Fixed Startup Cost (2020 $/MW) [d] |
|---|---|---|---|---|---|
| Flexible | 100% | 4 | 4 | 30 | 106 |
| Inflexible | 36% | 12 | 18 | 70 | 159 |

[a]. the ramping rates of CCGT can reach 2-8% $P_{nom}$/min, which means 5 mins from minimum load to maximum load. Our capacity expansion model only model demand at hourly resolution, thus treats ramping rates the same if they can ramp up/down from minimum load to maximum within 1 hr[18]. As for coal CCS, we are unable to find information regarding its ramping capability. However, Mai et al has evaluated the ramping rates of a demonstration CCGT-CCS plant and the tested ramp rates range from 0.6-5% per min. We chose the lower end of this range (0.6%/min or 36%/hr) as the ramping rates for the inflexible CCGT-CCS evaluated in this paper as a conservative proxy[19].
[b]. Data for minimum up/down time are adapted from the IRENA report[20] and the NREL report[21]. The data for Coal-CCS is adapted from a technical report for Petra Nova.[22]
[c]. The minimum turndown for CCGT is for a General Electric (GE) H class combined cycle.[6] The minimum turndown of Coal-CCS is from Petra Nova[22]. It should be noted that although it reported that the design operating ranges for Coal CCS is 50%-100%, the low-load operating point at Petra Nova was established at 70% as maintaining stable generation became difficult below this level.
[d] A 2012 NREL report (Table 1-1) provides the most detailed data on the startup costs and fuel consumption for different types of generators[17]. From their analysis, the median fixed cold startup costs for CCGT are 79 $/MW in 2011 USD, which is then converted to 2020 USD. The start-up costs associated with in flexible CCGT-CCS is assumed to be 150% of the flexible one.

To estimate the impacts of relevant policies on CCGT-CCS, we include two groups of policies: (1) $CO_2$ tax ranging from 50 to 200 $/t $CO_2$-equivalent ("Carbon tax"); and (2) clean energy standard (CES) ranging from 70 to 90%, coupled with 45Q $CO_2$ capture credits[23] ("Clean energy standard"). A CES requires a minimum share of electricity supply is produced by low- or zero-carbon generators, including wind, solar, nuclear, hydrogen combustion, and CCS equipped power plants. Each policy group is implemented both in the capacity expansion and unit commitment and economic dispatch modeling runs.

**Results and Discussions**
**Potential cycling patterns of CCGT-CCS under various $CO_2$ policies and CES requirements**
The 500 MW CCGT-CCS in ERCOT demonstrates different cycling patterns under various CES requirements, $CO_2$ policies, and flexibility levels (Figure 1). The number of start-ups of CCGT-CCS increases with higher CES requirement and $CO_2$ tax (Figure 1A). As more VRE capacity is added to ERCOT with more stringent CES requirement and higher $CO_2$ tax (Figure S1), it is economical to prioritize the use of wind and solar PV due to their zero marginal costs, leading to more frequent start up and shut down of thermal units to complement these weather dependent resources. As another result of increasing VRE, the capacity factor (average capacity utilization rate) for the CCGT-CCS plant decreases as the CES requirement and $CO_2$ tax increases (Figure 1B).

Although the overall trend is similar between "Carbon tax" and "Clean energy standard" as the regulation gets stringent, the choice of $CO_2$ policy has a strong impact on the start-up number and the utilization level of CCS. Specifically, CCGT-CCS cycles more frequently in scenarios with a $CO_2$ tax compared to those with $CO_2$ subsidies (56 – 76 annual start-ups under "Carbon tax" and 3 – 48 annual start-ups under "Clean energy standard"). This difference is particularly evident at less stringent CES requirements. In such cases, scenarios with $CO_2$ tax have over 50 start-ups/year, whereas those with $CO_2$ subsidies have less than 6 start-ups. This is because $CO_2$ taxes increase the cost of thermal generators that emit $CO_2$, leading to higher and more volatile electricity prices, and thus more frequent start-ups; $CO_2$ subsidies, in contrast, reduce the generation cost for CCGT-CCS, improving its position in the order of least-cost dispatch and resulting in more consistent operations. For the same reason, we also observe higher capacity factors of CCGT-CCS when CCS tax credits are available than those with carbon taxes (0.37 – 0.60 under "Carbon tax" and 0.44 – 0.86 under "Clean energy standard").

The operational behaviors for CCGT-CCS between "Carbon tax" and "Clean energy requirement" are also different (Figures 1C & 1D). During the period when marginal electricity prices are low, we find CCGT-CCS under $CO_2$ tax tend to be switched off to avoid the loss of profits as the marginal generation cost of plant is around $30/MWh. However, with $CO_2$ credits that subsidizes $CO_2$ capture and sequestration and reduces the plant's marginal generation cost, CCGT-CCS is more likely to operate at the minimum load to avoid cycling costs. After accounting for $CO_2$ transport & storage costs, the $CO_2$ capture subsidy of $85/t $CO_2$ is effectively a production subsidy equivalent to $26/MWh for the CCGT-CCS unit, reducing its marginal cost to $5/MWh, which increases the number of hours when CCGT-CCS is economic to dispatch and reduces incentives to cycle off/on to avoid short periods of lower prices (Figure 1C). However, when the CES requirement increases to 90%, negative electricity prices occur more frequently during periods of renewable energy curtailment, and the difference between electricity prices and CCGT-CCS generation cost becomes much larger, even for $CO_2$ subsidy scenarios, resulting in more frequent start up and shut down to avoid operating at a loss during these periods Figure 1D).

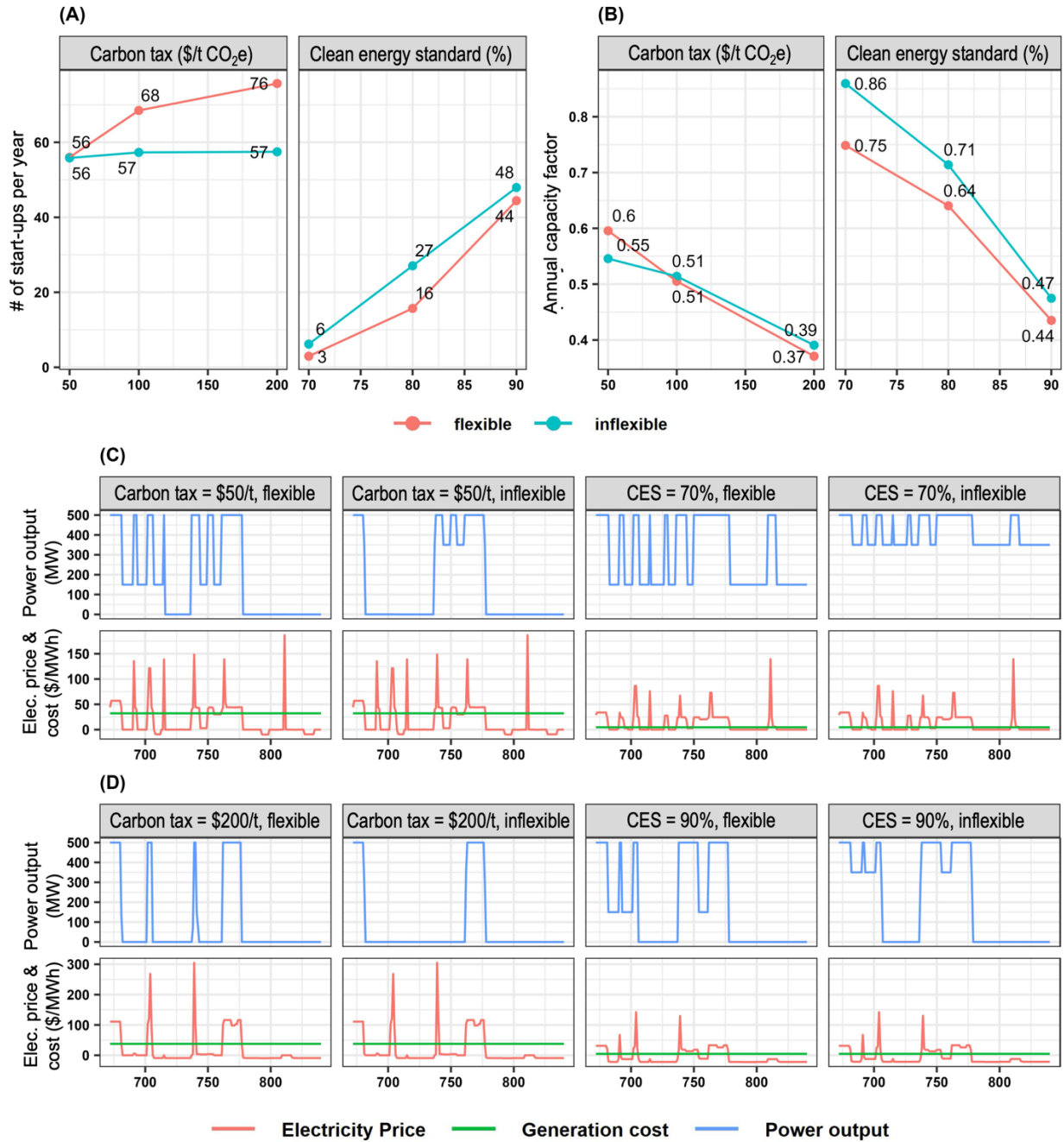

**Figure 1.** (A) The number of startup events and (B) capacity factor for a 500 MW CCGT-CCS plant at varying CES requirements and $CO_2$ policies. The power output (blue line), electricity price (red line), and generation cost of CCGT-CCS (green line) under (C) CCS credits + 70% CES and carbon tax of 50$/t $CO_2$ and (D) 90% CES requirements and carbon tax of 200 $/t $CO_2$ during the first week of February.

In addition to the impacts from policies, varying flexibility levels also lead to different cycling behaviors for CCGT-CCS across various scenarios. Notably, flexible CCGT-CCS exhibits a higher numbers of start-ups in $CO_2$ tax scenarios compared to its inflexible counterparts when carbon taxes are high ($100 and $200/ton $CO_2$e). A $CO_2$ tax leads to greater electricity price volatility, which tends to benefit CCGT-CCS with more flexible technical features that can perform more frequent cycling, producing electricity only when it is most needed, and avoid low-price periods when the plant would otherwise operate at a loss. On the other hand, in scenarios with CCS credits, flexible CCGT-CCS experiences fewer start-ups than inflexible CCGT-CCS due to relatively lower electricity price volatility. During times when electricity price is zero or negative, flexible CCGT-CCS can operate at a lower stable power generation level. This behavior avoids the costs associated with cycling and minimizes operating losses during these periods. In contrast, inflexible CCGT-CCS, with a minimum stable power generation assumed to be 70%, will incur higher operating losses during periods of negative electricity prices, making it more economical for inflexible CCGT-CCS to shut down and avoid power generation during these periods. These results suggest that under policies penalizing $CO_2$ emissions ($CO_2$ taxes or emissions cap), CCGT-CCS might be designed to either achieve lower minimum stable generation level *or* to manage regular startup and shutdown events (e.g., weekly start-ups). In the context of a subsidy for $CO_2$ capture and storage, CCGT-CCS plants may start-up less frequently, while being able to maintain a lower minimum stable generation level can effectively reduce the needs for frequent start up and shut down in this policy environment.

**Increased profits for plant owners from more flexible CCGT-CCS**

We find that CCGT-CCS with enhanced flexibility exhibit higher operating profits (e.g., excluding fixed operations and maintenance and repayment of capital) in all the evaluated scenarios, compared to the inflexible CCGT-CCS (see Figure 2A). With carbon taxes, improved flexibility leads to $7 million to $8 million more operational profits per year, and this accounts for around 20 - 25% of the total annual fixed operations and maintenance costs for a 500 MW CCGT-CCS plant. However, enhanced flexibility of CCGT-CCS is less valuable in the context of $CO_2$ subsidies than in $CO_2$ tax scenarios. With tax credits for $CO_2$ capture and storage, improved flexibility increases profits by only $3 million to $4 million, with profits increasing with the stringency of clean energy standards.

We further explore to what extent each flexibility parameter benefits the operating profits by focusing on scenarios with CCS credits + 90% CES and 200 $/t $CO_2$, respectively (Figure 2B). Starting from the most inflexible CCGT-CCS, we improve one flexibility metric in each unit commitment and economic dispatch run. Under both carbon policy cases, we find that a lower minimum power level will result in the highest profits, followed by lower start-up costs. With a carbon tax of $200/ton, faster ramping and lower up/down time can increase the profits of CCGT-CCS by only $0.7 – $0.9 million per year, respectively; and they have almost no impacts on the operational profits when a 90% CES is implemented. Shorter minimum up/down time even slightly reduces the profit of CCGT-CCS because this improved flexibility leads to fewer running hours of CCGT-CCS. The limited impacts from minimum up/down time can be explained by the fact that the CCGT-CCS plant tend to stay on or off for long hours even with shorter minimum up/down time requirements (Figure S2), suggesting that the minimum up/down time constraints are not binding in most scenarios.

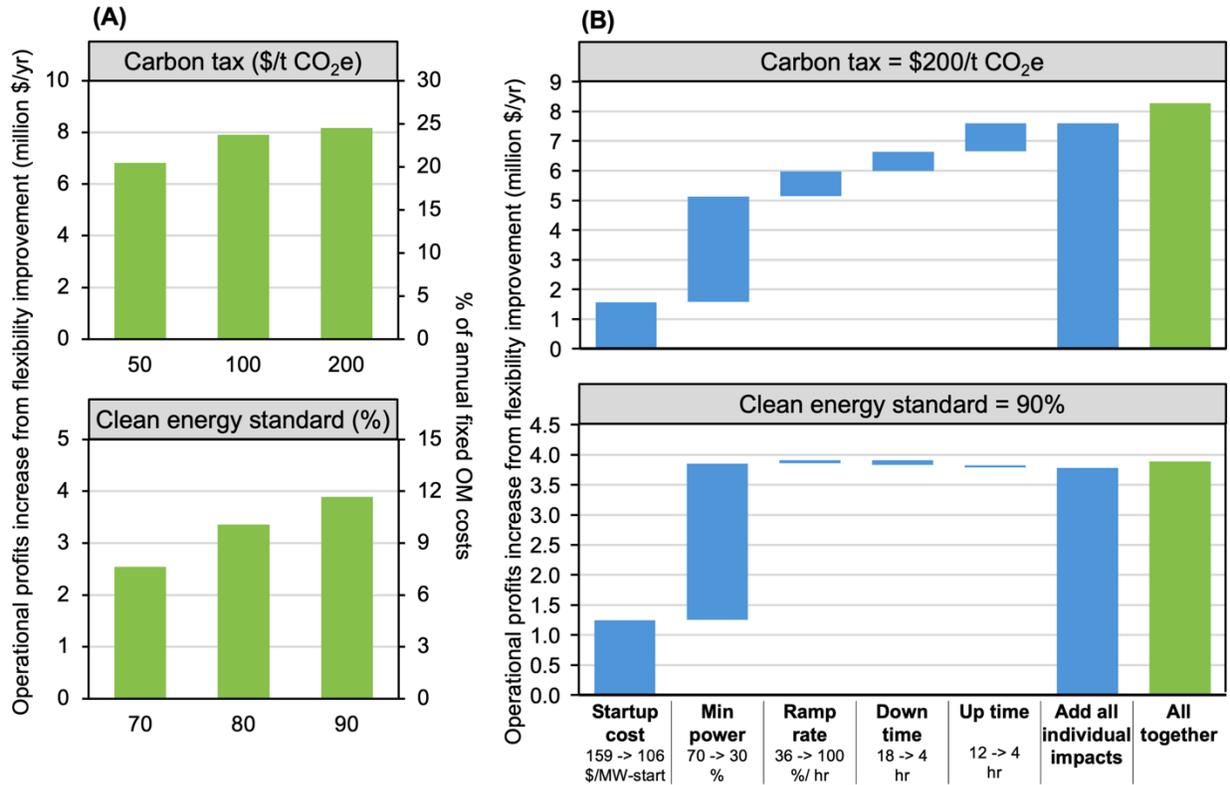

**Figure 2.** (A). The increase in operating profits for CCGT-CCS from improving all flexibility parameters all at once. (B) The increase in operating profits carbon tax of $200/t $CO_2$ and CES requirement of 90% (with 45Q) when each flexibility parameter independently changes from the most inflexible to the most flexible. "Add all individual impacts" is the total impacts of each flexibility parameter improvement (assuming no interaction among each parameter) and "all together" is the increased operating profits where all flexibility parameters are included at once (including interaction among each parameter).

However, by looking at the parameters one by one, this approach does not capture the potential interactions among these parameters. Indeed, we observe that the profits from a case where all parameters are improved all at once (results in Figure 2A or "All together" in Figure 2B) are higher than the sum of increased profits from each individual parameter improvement ("Add all individual impacts" in Figure 2B), indicating the existence of interactions among these flexibility parameters, especially when there is a carbon tax. Therefore, we run additional operational cases with different combinations of flexible parameters, including two, three, and four improved parameters in each run (Tables S2 and S3). Under both policy scenarios, the existence of a lower minimum power level would significantly increase the additional profits from faster ramping (from $0.9 million to $2.5 million under the carbon tax scenario and from $0.05 million to $0.34 million under the 90% CES scenario). With deeper turndown levels, fast ramping permits CCGT-CCS plants to ramp from max to min output and vice versa in an hour or less, while slower ramping rates may constrain the ability to turn down/up to maximize operation profits. Similarly, lower startup costs are more valuable in combination with shorter up/down time, which together enable more frequent cycles. However, we also observe benefits canceling out in some cases. For example, with lower minimum power levels, the benefits from shorter up/down time would be significantly reduced, especially when carbon tax is applied (reduced from $ 0.7 - 1 million to $ 0 - 0.1 million), indicating that it might not be necessary to improve all flexibility parameters all at once. These results emphasize the importance of considering the interactions among flexibility parameters when estimating the tradeoff between the extra costs to improve the flexibility and the additional operational profits of CCGT-CCS.

**The value of improved operating flexibility for the power system**

By investigating the capacity expansion results with CCGT-CCS, we further explore how enhanced operating flexibility might affect the market share of CCGT-CCS, integration of renewable resources (VRE), and total system cost. When a carbon tax of $200/ton is applied, improving all flexibility parameters can increase the capacity of CCGT-CCS from 2.1 to 8.0 GW ("All together" in Figure 3A). However, a lower minimum power level alone can increase the capacity of CCGT-CCS to 6.7 GW, as the marginal value of other flexibility parameters is substantially lower if plants can achieve a lower minimum power level (Table S4). For the same reason, total capacity of new renewable resources added to ERCOT increases by 1.7% if all parameters are improved and 1.6% if only the minimum power level is lower. However, interactions among all parameters will lead to greater system cost reductions ("All together" in Figure 3A). For example, the combination of lower minimum power and faster ramping reduces renewable curtailment, resulting in less variable costs and carbon taxes incurred. In the scenario with 90% CES + 45Q CCS credits, we observe that 6.6 GW of CCGT-CCS is added to ERCOT even without any flexibility improvement, but the overall enhanced system flexibility only increases the capacity by 13% to 7.4 GW ("All together" in Figure 3B). This increase in CCGT-CCS deployment is driven solely by a lower minimum power level, which alone can also increase the capacity of CCGT-CCS to 7.4 GW. Instead of promoting the addition of new renewable resources, more flexible CCGT-CCS reduces renewable energy capacity by 2.6% (from 115 GW to 112 GW) due to reduced renewable energy curtailment. Under both policy scenarios, we find that the economic benefits from greater operating flexibility at the system level are modest: enhanced flexibility in CCGT-CCS reduces total electricity system costs by only 0.5% and 0.3 % in the carbon tax of $200/t $CO_2$e and 90% CES with $CO_2$ credit scenarios, respectively. These observations are consistent with results from Pratama et al. (2022), which demonstrates the benefit from reducing the minimum stable power output of a CCGT-CCS is negligible.[9]

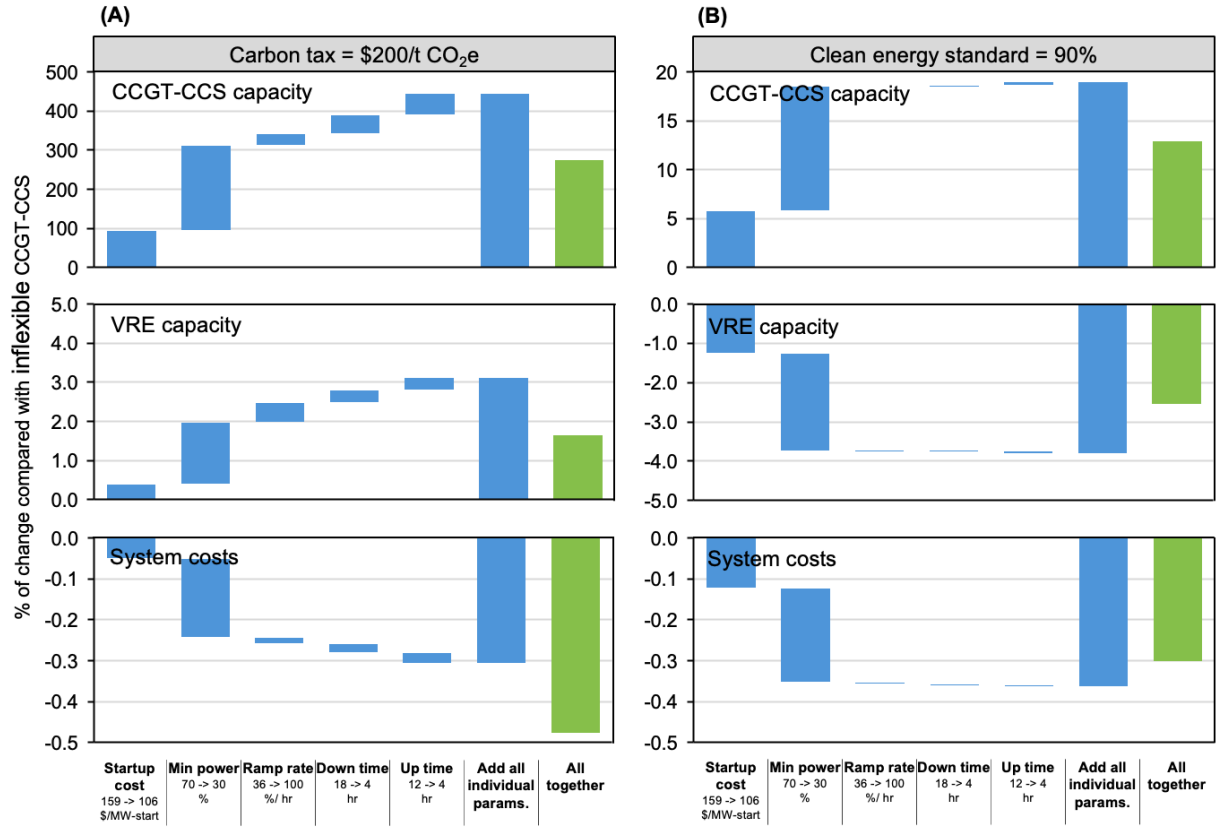

**Figure 3.** The change in CCGT-CCS capacity deployment (top), new renewable capacity (middle), and total system cost (bottom) at carbon tax of $200/t $CO_2$ and CES requirement of 90% (with 45Q) when each flexibility parameter independently changes from the most inflexible to the most flexible. "Add all individual impacts" is the total impacts of each flexibility parameter improvement (assuming no interaction among each parameter) and "all together" is the increased operating profits where all flexibility parameters are included at once (including interaction among each parameter).

**Conclusions**

In summary, we use an open-source capacity expansion model with unit commitment and economic dispatch constraints to investigate the value of improved operating flexibility of a combined cycle gas turbine with carbon capture and sequestration (CCGT-CCS). We assess the impacts of faster ramping rates, lower stable power generation, shorter minimum up/down time, and lower start-up costs on plant-level profitability, total deployed capacity, and electricity system costs. By comparing two policy scenarios (i.e., carbon tax and CCS credits + clean electricity standard), our analysis also highlight the importance of policy environment in determining the value of different flexibility improvement: e.g., the importance of frequent startup and shutdown when electricity prices are volatile versus maintaining a lower minimum stable generation level when CCS credits are available.

We find that enhanced flexibility of CCGT-CCS plants is beneficial for both plant owners and total system costs, though the impacts are relatively modest. For a 500 MW CCGT-CCS plant in ERCOT, the difference between the most flexible and least flexible characteristics assessed herein represents an increase in annual operating profits by $6.9 - 8.3 million under carbon tax and $2.6 - 3.9 million under $CO_2$ capture subsidy scenarios, equivalent to 8 - 25% of the plant's annual fixed O&M costs. We find the most value when plants are designed either for frequent cycling—with lower startup costs and, secondarily, shorter minimum up/down times—*or* for lower turn-down—with lower minimum stable power levels and, secondarily, faster ramping rates. Improving both cycling and turn-down flexibility appears unnecessary.

Additionally, improved operating flexibility can increase the market penetration of CCGT-CCS plants, with installed capacity increasing from 6.6 to 7.4 GW in the scenario with 90% CES with 45Q and from 2.1 to 8.0 GW with a carbon tax of $200/t $CO_2$e. However, we find that flexibility only results in modest reductions in total electricity system costs (by 0.3% and 0.5%, under the two scenarios, respectively).

Although we find that enhanced operating flexibility increases profitability of CCGT-CCS plants and reduces overall system costs, increasing operating flexibility generally comes with additional costs, which are not considered in this study. Also, additional measures are needed to improve the performance of $CO_2$ capture during frequent startup and shutdown events, which may lead to additional investment[24]. As the benefits of operating flexibility from this study are relatively modest, enhanced flexibility of CCGT-CCS plants is only warranted if it entails a limited increase in capital and maintenance costs. Instead, reductions in overall capital costs are likely to be most determinative of future deployment and system-level impacts of gas power plants with CCS.

**Acknowledgement:**
Funding for this work was provided by the U.S. Department of Energy Advanced Research Projects Agency-Energy (ARPA-E) through grant award DE-AR0001285.

# Supplementary information

**Table S1.** The key cost assumptions for different technologies used in this study (from NREL Annual Technology Baseline 2022).

| Technology | CapEx ($/kW) | CapEx ($/kWh) | Fixed O&M costs ($/kW-yr) | Fixed O&M costs ($/kWh-yr) | Variable O&M costs ($/MWh) |
|---|---|---|---|---|---|
| Natural gas combined cycle | 920 | - | 28 | - | 2 |
| Natural gas combustion turbine | 793 | - | 21 | - | 5 |
| Natural gas combined cycle with carbon capture and storage | 2,310 | - | 67 | - | 10 |
| Battery storage | 159 | 114 | 6 | 5 | 0 |
| Nuclear | 4,388 | - | 146 | - | 3 |
| Land-based wind | 1,053 | - | 39 | - | 0 |
| Utility PV | 845 | - | 16 | - | 0 |
| Offshore wind | 1,722 | - | 89 | - | 0 |

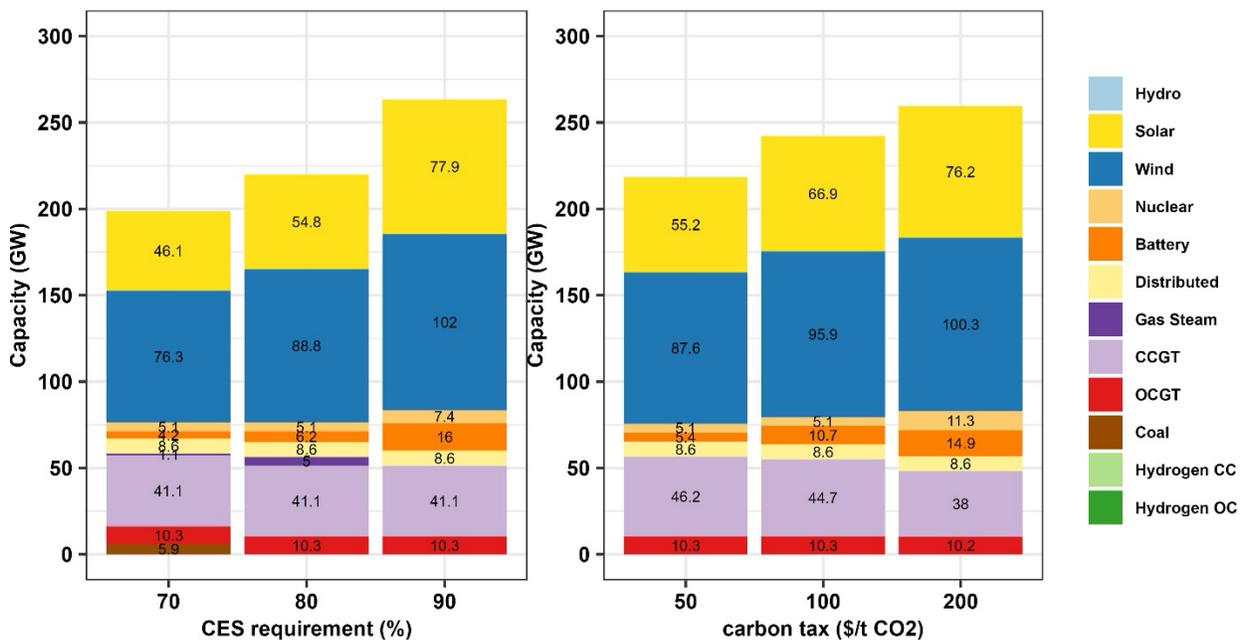

**Figure S1.** Least cost capacity mixes at ECROT in 2035 with CES requirements from 70% to 90% and CO2 tax from 50 to 200 $/t CO2 (CCGT-CCS is purposely not included for the least cost expansion).

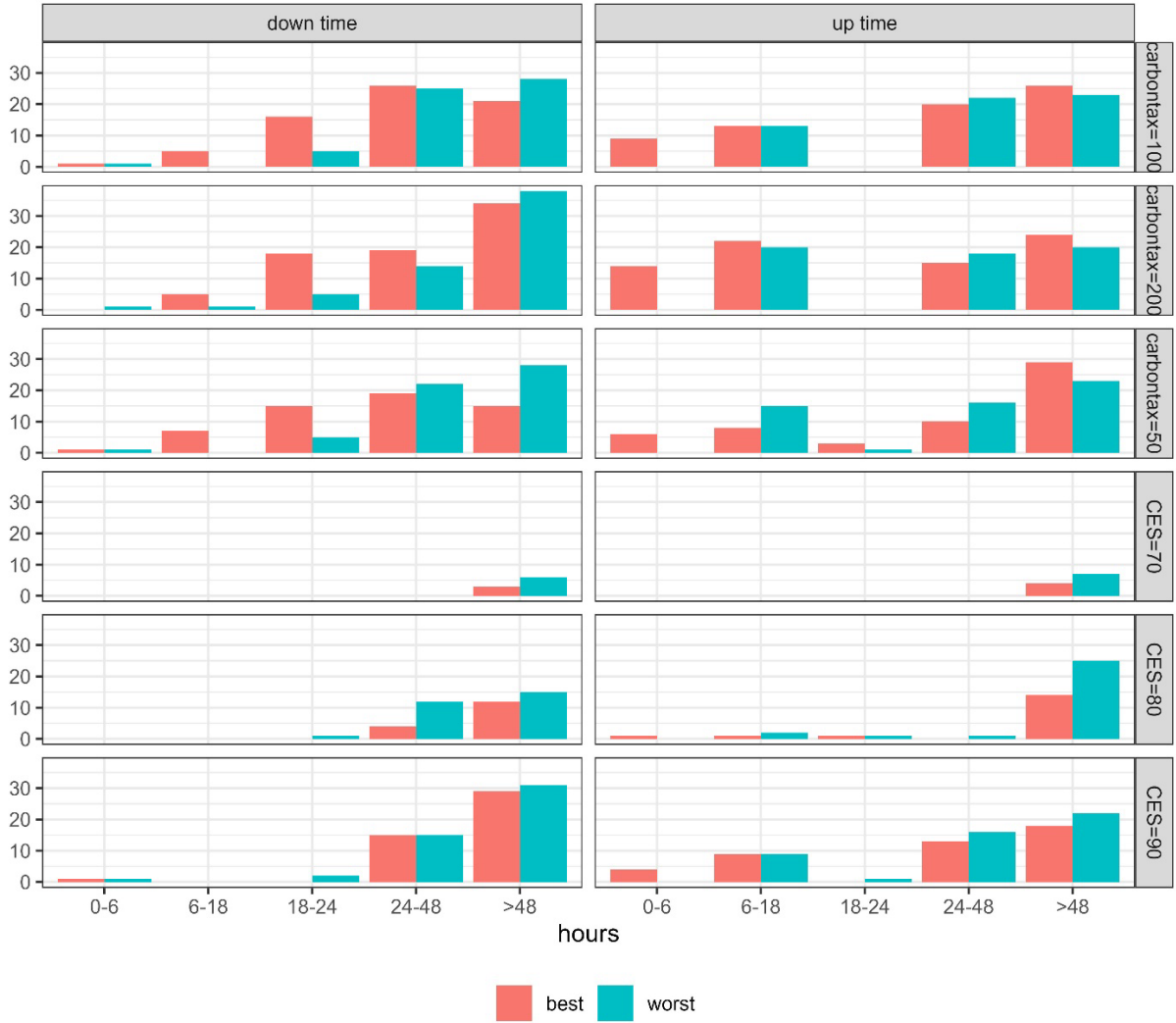

**Figure S2.** The distribution of the up time and downtime for CCGT-CCS at various CES requirements and CO2 policies.

**Table S2.** The additional profits (million $) from improving the flexibility parameters in a 500 MW CCGT-CCS power plant in ERCOT when a carbon tax of $200/ton $CO_2e$ is applied. Compared with the case where no flexibility improvements are included. Different combinations of flexibility parameter improvements are included.

|    | **One** | **Two** | | | | | **Three** | | | | | | **Four** | | | | **All** |
|----|---------|---------|----|----|----|----|-----------|----|----|----|----|----|----------|----|----|----|---------|
|    | None | P1 | P2 | P3 | P4 | P5 | P1+P2 | P1+P3 | P1+P4 | P2+P3 | P2+P4 | P3+P4 | P1+P2+P3 | P1+P2+P4 | P1+P3+P4 | P2+P3+P4 | P1+P2+P3+P4 |
| P1 | 1.6 | - | - | - | - | - | - | - | - | - | - | - | - | - | - | - | - |
| P2 | 3.6 | 5.1 | - | - | - | - | - | - | - | - | - | - | - | - | - | - | - |
| P3 | 0.9 | 2.5 | 6.1 | - | - | - | 7.7 | - | - | - | - | - | - | - | - | - | - |
| P4 | 0.7 | 2.9 | 3.6 | 1.9 | - | - | 5.1 | 4.5 | - | 6.1 | - | - | 7.7 | - | - | - | - |
| P5 | 1.0 | 2.9 | 3.7 | 2.0 | 1.7 | - | 5.5 | 4.0 | 4.6 | 6.3 | 3.7 | 3.2 | 8.2 | 5.5 | 6.4 | 6.3 | 8.3 |

Where,

P1: startup costs from $159 to $106 per MW per start
P2: minimum power level from 70% to 30%
P3: ramp rate from 36% to 100% of capacity per hour
P4: minimum down time from 18 hours to 4 hours
P5: minimum up time from 12 hours to 4 hours

**Table S3.** The additional profits (million $) from improving the flexibility parameters in a 500 MW CCGT-CCS power plant in ERCOT when a clean energy standard at 90% and 45Q tax credits for CCS are applied. Compared with the case where no flexibility improvements are included. Different combinations of flexibility parameter improvements are included.

|    | One | Two | | | | | Three | | | | | | Four | | | | All |
|----|-----|-----|---|---|---|---|-------|---|---|---|---|---|------|---|---|---|-----|
|    | None | P1 | P2 | P3 | P4 | P5 | P1+P2 | P1+P3 | P1+P4 | P2+P3 | P2+P4 | P3+P4 | P1+P2+P3 | P1+P2+P4 | P1+P3+P4 | P2+P3+P4 | P1+P2+P3+P4 |
| P1 | 1.3 | - | - | - | - | - | - | - | - | - | - | - | - | - | - | - | - |
| P2 | 2.6 | 3.5 | - | - | - | - | - | - | - | - | - | - | - | - | - | - | - |
| P3 | 0.1 | 1.5 | 2.9 | - | - | - | 3.9 | - | - | - | - | - | - | - | - | - | - |
| P4 | -0.1 | 1.3 | 2.6 | 0.1 | - | - | 3.5 | 1.4 | - | 3.0 | - | - | 3.9 | - | - | - | - |
| P5 | 0 | 1.3 | 2.6 | 0.1 | -0.1 | - | 3.5 | 1.5 | 1.4 | 3.0 | 2.6 | 0.1 | 3.9 | 3.5 | 1.5 | 3.0 | 3.9 |

Where,
P1: startup costs from $159 to $106 per MW per start
P2: minimum power level from 70% to 30%
P3: ramp rate from 36% to 100% of capacity per hour
P4: minimum down time from 18 hours to 4 hours
P5: minimum up time from 12 hours to 4 hours

**Table S4.** The additional installed capacity (GW) of CCGT-CCS power plant in ERCOT when a carbon tax of $200/ton $CO_2e$ is applied. Compared with the case where no flexibility improvements are included. Different combinations of flexibility parameter improvements are included.

|    | One | Two | | | | | Three | | | | | | Four | | | | All |
|----|-----|-----|---|---|---|---|-------|---|---|---|---|---|------|---|---|---|-----|
|    | None | P1 | P2 | P3 | P4 | P5 | P1+P2 | P1+P3 | P1+P4 | P2+P3 | P2+P4 | P3+P4 | P1+P2+P3 | P1+P2+P4 | P1+P3+P4 | P2+P3+P4 | P1+P2+P3+P4 |
| P1 | 2.0 | - | - | - | - | - | - | - | - | - | - | - | - | - | - | - | - |
| P2 | 4.6 | 5.2 | - | - | - | - | - | - | - | - | - | - | - | - | - | - | - |
| P3 | 0.6 | 2.8 | 5.0 | - | - | - | 5.8 | - | - | - | - | - | - | - | - | - | - |
| P4 | 2.0 | 3.8 | 4.6 | 2.0 | - | - | 5.2 | 4.4 | - | 5.0 | - | - | 5.8 | - | - | - | - |
| P5 | 1.2 | 3.3 | 4.7 | 1.4 | 1.9 | - | 5.3 | 3.5 | 4.5 | 5.1 | 4.7 | 2.8 | 5.8 | 5.4 | 4.7 | 5.0 | 5.8 |

Where,
P1: startup costs from $159 to $106 per MW per start
P2: minimum power level from 70% to 30%
P3: ramp rate from 36% to 100% of capacity per hour
P4: minimum down time from 18 hours to 4 hours
P5: minimum up time from 12 hours to 4 hours

**Table S5.** The additional installed capacity (GW) of CCGT-CCS power plant in ERCOT when a clean energy standard at 90% and 45Q tax credits for CCS are applied. Compared with the case where no flexibility improvements are included. Different combinations of flexibility parameter improvements are included.

|    | One  | Two |     |     |     |     | Three |     |     |     |     |     | Four |     |     |     | All |
| -- | ---- | --- | --- | --- | --- | --- | ----- | --- | --- | --- | --- | --- | ---- | --- | --- | --- | --- |
|    | None | P1  | P2  | P3  | P4  | P5  | P1+P2 | P1+P3 | P1+P4 | P2+P3 | P2+P4 | P3+P4 | P1+P2+P3 | P1+P2+P4 | P1+P3+P4 | P2+P3+P4 | P1+P2+P3+P4 |
| P1 | 0.4  | -   | -   | -   | -   | -   | -     | -   | -   | -   | -   | -   | -    | -   | -   | -   | -   |
| P2 | 0.8  | 0.9 | -   | -   | -   | -   | -     | -   | -   | -   | -   | -   | -    | -   | -   | -   | -   |
| P3 | 0    | 0.4 | 0.8 | -   | -   | -   | 0.9   | -   | -   | -   | -   | -   | -    | -   | -   | -   | -   |
| P4 | 0    | 0.4 | 0.8 | 0   | -   | -   | 0.9   | 0.4 | -   | 0.8 | -   | -   | 0.9  | -   | -   | -   | -   |
| P5 | 0    | 0.4 | 0.8 | 0   | 0   | -   | 0.9   | 0.4 | 0.4 | 0.8 | 0.8 | 0   | 0.9  | 0.9 | 0.4 | 0.8 | 0.9 |

Where,
P1: startup costs from $159 to $106 per MW per start
P2: minimum power level from 70% to 30%
P3: ramp rate from 36% to 100% of capacity per hour
P4: minimum down time from 18 hours to 4 hours
P5: minimum up time from 12 hours to 4 hours